\documentclass[11pt,a4paper]{article}
\pdfoutput=1 

\usepackage{jheppub-nofpheader}
\usepackage{latexsym}
\usepackage{revsymb}

\usepackage{bm}
\usepackage{slashed}

\gdef\@fpheader{\@empty}

\newcommand{\Lstar}{L_*}
\newcommand{\Tstar}{T_*}
\newcommand{\taustar}{\tau_*}
\newcommand{\tritium}{{}^3_1\mathrm{H}}
\newcommand{\HeThree}{{}^3_2\mathrm{He}}
\newcommand{\Dy}{{}^{163}_{\phantom{0}66}\mathrm{Dy}}
\newcommand{\Ho}{{}^{163}_{\phantom{0}67}\mathrm{Ho}}
\newcommand{\Co}{{}^{57}_{27}\mathrm{Co}}
\newcommand{\Fe}{{}^{57}_{26}\mathrm{Fe}}
\newcommand{\bra}[1]{\langle #1 |}
\newcommand{\ket}[1]{| #1 \rangle}

\preprint{IPMU12-0189}

\title{Using Neutrinos to test the Time-Energy Uncertainty Relation in an Extreme Regime}

\author[a]{Ramaswamy S. Raghavan,}
\author[a]{Djordje Minic,}
\author[a,b]{Tatsu Takeuchi,}
\author[a]{and Chia Hsiung Tze}

\affiliation[a]{Center for Neutrino Physics, Department of Physics, Virginia Tech, Blacksburg, VA 24061, USA} 
\affiliation[b]{Kavli Institute for the Physics and Mathematics of the Universe,
University of Tokyo, Kashiwa-shi, Chiba-ken 277-8568, Japan}

\emailAdd{dminic@vt.edu}
\emailAdd{takeuchi@vt.edu}
\emailAdd{kahong@vt.edu}

\abstract{
We discuss a direct test of the relation of time and energy in the very long-lived decay of tritium ($\tritium$) (meanlife $\tau\sim 18\,\mathrm{yrs}$) with the width $\Gamma\sim 10^{-24}\,\mathrm{eV}$ [set by the time-energy uncertainty (TEU)], using the newfound possibility of resonance reactions 
$\tritium\leftrightarrow\HeThree$ with $\Delta E/E\sim 5\times 10^{-29}$.  
The TEU is a keystone of quantum mechanics, but probed for the first time in this extreme time-energy regime. Forestalling an apparent deviation from the TEU, we discuss the ramifications and a possible generalization of the TEU as $\Delta E\,\Delta t \ge (\hbar/2)[1+(\Delta t/\Tstar)^n]$ where $\Delta t = \tau$ is the time of measurement (the lifetime of the state), $\Tstar=\Lstar/c$ the time for light to cross the Universe $\sim 3\times 10^{18}\,\mathrm{s}$, and $n$ a parameter subject to future measurements.
(by R.~S.~Raghavan.)
}

\arxivnumber{1210.XXXX}

\dedicated{\vspace{1cm}
{\normalsize This paper is dedicated to the memory of Raju Raghavan,\\
a dear colleague and a wonderfully creative physicist.\\}
\vspace{3cm}
}

\notoc
\begin{document}
\maketitle
\flushbottom

\section{Introduction}

In this note, we collect some essential thoughts from an unfinished collaboration with our dear colleague Ramaswamy (Raju) S. Raghavan, concerning the role of neutrino probes in the foundations of quantum theory and in the search for the elusive quantum theory of gravity. This project was interrupted by Raju's untimely passing on October 20, 2011.
The current paper is motivated by the occasion of the Virginia Tech Symposium on the Life and Science of Dr. Raju Raghavan (October 20, 2012).

\bigskip
The starting point of our collaboration was a proposal by Raghavan \cite{Raghavan:2005gn,Raghavan:2006xf,Raghavan:2009hj,Raghavan:2009dk,Raghavan:2009zz,Raghavan:2009sf} 
to use M\"ossbauer neutrinos to check the time-energy uncertainty (TEU) relation
$\Delta t\,\Delta E\ge \hbar/2$ in the regime of extremely large uncertainty in time: $\Delta t \sim 10^9\,\mathrm{s}$, and correspondingly small uncertainty in energy:
$\Delta E\sim \hbar/\Delta t\sim 10^{-43}\,\mathrm{J}\sim 10^{-24}\,\mathrm{eV}$.
His idea was to perform a neutrino version of the famous experiment by Wu et al. from 1960 \cite{wu:1960,lynch:1960,harris:1960} which utilized the then recently discovered M\"ossbauer effect in $\gamma$-rays \cite{Mossbauer:1958aa,Mossbauer:1958bb,Mossbauer:1959,Mossbauer:1962xd}.

It had been suggested by Mead (1966) \cite{Mead:1964zz,Mead:1966zz} that 
possible manifestations of a fundamental length scale, such as the Planck scale
$\ell_P = \sqrt{G\hbar/c^3}$, could be seen in this 
small $\Delta E$ regime
as deviations of $\Delta E$ from the expected $\sim\hbar/\Delta t$ behavior.
We will discuss this claim in more detail later.
Following Mead, Andryushin and Melnikov (1973) \cite{Andryushin:1973qk}
proposed the idea of
using the M\"ossbauer effect to measure the width of the resonant reaction 
$\tritium \leftrightarrow \HeThree + \bar{\nu}_e$ to extreme accuracy to look for
such a deviation.
Kells and Schiffer (1984) \cite{Kells:1981xt,Kells:1982rm,Kells:1984nm} discussed 
the necessary conditions for such an experiment to be implemented. 
To this, Raghavan brought in the bold proposal of using Niobium tritide as the actual source of recoilless emission and absorption of the neutrinos.

Raghavan's proposal lead to a couple of intense controversies, one on whether the
M\"ossbauer effect could truly be realized using Niobium tritide \cite{Potzel:2009qe,Potzel:2009pr,Potzel:2011zza,Schiffer:2009zz,Raghavan:2009zz,Suzuki:2010zza},
and another on the interpretation of the Time-Energy uncertainty relation 
and its implications for the proposed experiments \cite{Lipkin:2009uq,Bilenky:2008ez,Bilenky:2008dk,Bilenky:2009zz,Bilenky:2011pk,Akhmedov:2008jn,Akhmedov:2008zz,Kopp:2009fa}.
At the same time, it generated a great deal of excitement and interest on the 
new kinds of experiments that would be made possible if M\"ossbauer neutrinos became
a reality, and the types of physics it will allow us to probe \cite{Minakata:2006ne,Minakata:2007tn,Parke:2008cz,Machado:2011tn}.
For instance, it would be possible to perform neutrino oscillation experiments 
on a table-top scale \cite{Raghavan:2005gn,Raghavan:2006xf}.
We could also redo all the M\"ossbauer effect based
$\gamma$-ray experiments with neutrinos, providing us with a non-electromagnetic
probe into nature's workings.

The three of us (Minic, Takeuchi, Tze) came into the picture as theorists
interested in understanding quantum gravity, and how the existence of a
fundamental length scale may manifest itself in nature.
The task Raghavan charged us with was: 
in the event that Raghavan's experiment was performed and a deviation from
the $\Delta E\sim \hbar/\Delta t$ behavior is discovered, what would it tell us
about the nature of space-time and quantum gravity?  
In this paper, we would like to present some of our thoughts on this subject.

But before we do that, we would like to begin with some of the background
and arguments surrounding Raghavan's proposal, and what all the fuss was about.
This, we do in section 2 where we review the history and controversies 
on the idea of M\"ossbauer neutrinos, and those on the Time-Energy uncertainty relation.
Section 3 is based on a draft written by Raghavan dated March 23, 2010,
in preparation for a paper the four of us would have written together.
In it, he discusses his main ideas and claims on what his experiment could
have achieved.
We have retained his original words as much as possible, except for a few
corrections of typographical errors and changes of some symbols for notational consistency.
Footnotes have been added where we felt some clarification was called for.
Section 4 is based on our contributions during the many meetings the three of us had with Raghavan between Fall 2009 and Spring 2011, in which he poked our brains for 
possible theoretical ways that the canonical time-energy uncertainty relation could
be violated.
In section 5, we conclude with some comments on Raghavan's legacy.


\section{Background}

\subsection{M\"ossbauer Neutrinos}

The idea of enhancing the neutrino detection cross section via resonant 
recoilless neutrino emission and absorption, \textit{i.e.} the M\"ossbauer effect (ME),
is not new.
Almost immediately after the discovery of the ME in $\gamma$-rays (1958) \cite{Mossbauer:1958aa,Mossbauer:1958bb,Mossbauer:1959,Mossbauer:1962xd}, 
Visscher proposed to use monochromatic neutrinos from
an electron capture process to achieve a similar effect (1959) \cite{Visscher:1959}.\footnote{%
Visscher in Ref.~\cite{Visscher:1959} credits D.~Nagle for the original idea of
using the ME for neutrino detection.}
The idea was further developed by Kells and Schiffer in the early 1980's \cite{Kells:1981xt,Kells:1982rm,Kells:1984nm}
where they also considered the possibility of obtaining
monochromatic (anti)neutrinos from bound-state $\beta$-decay.

Bound-state $\beta$-decay refers to the process in which the electron emitted
by a $\beta$-decaying nucleus is captured by the daughter nucleus instead
of being emitted into an unbound continuum state.
For instance, the electron from tritium $\beta$-decay could be captured
in the $K$-shell of the daughter $\HeThree$ nucleus leading effectively to a two-body decay:
\begin{equation}
\tritium\,(1s) \quad\rightarrow\quad 
\HeThree\,(1s^2) + \bar{\nu}_e\,(18.6\,\mathrm{keV})\;.
\label{bbtritium}
\end{equation}
Originally proposed as a possible $\beta$-decay process by Daudel et al. in 1947 \cite{Daudel:1947aa,Daudel:1947bb,Daudel:1947cc}, the theory of
bound-state $\beta$-decay was developed 
later by Sherk (1959) \cite{Sherk:1959} and Bahcall (1961) \cite{Bahcall:1961zz}.
Bahcall calculated the ratio of bound to continuum decay rates in
atomic tritium and found
\begin{equation}
\Gamma_\mathrm{Bound}/\Gamma_\mathrm{Continuum} \;=\; 0.0069 \;=\; 0.69\,\%\;,
\end{equation}
of which 22\% was attributed to electron absorption by excited states.
So the fraction to the atomic ground state of $\HeThree$
is $0.54\%$,
which is the number cited by Raghavan in Refs.~\cite{Raghavan:2006xf,Raghavan:2009hj}.

It should be noted that the above bound-state $\beta$-decay process for tritium
has never been observed experimentally.  
This is not surprising given the difficulty in observing the emitted $\bar{\nu}_e$,
which is what Raghavan was trying to achieve via the ME.
It has been suggested that one may be able to see this
process in an ion accelerator \cite{Takahashi:1987zz}, 
but as far as we are aware, the experiment is yet to be carried out.
In fact, the only experimentally detected bound-state $\beta$-decay process
so far is that of $\Dy^{66+}$ \cite{Jung:1992pw}.

The only evidence that bound-state $\beta$-decay exists for tritium is indirect:
Budick (1983) \cite{Budick:1983} noted that the various measurements of the tritium lifetime
differed systematically depending on the environment that the decaying tritium was placed in.
Since both the bound-state and continuum decay rates of tritium depend on the
electronic environment of the nucleus \cite{Bahcall:1961zz,Bahcall:1963zza}, 
Budick calculated the decay rates
of atomic $\tritium$, the ions $\tritium^+$ and $\tritium^-$, and molecular $\tritium$
and found that the continuum rates differed little 
while the differences in the bound-state rates led to a
qualitative agreement with the different lifetime measurements.\footnote{%
Sherk had already noted that the lifetimes of atomic
and molecular tritium should be different in Ref.~\cite{Sherk:1959} (1959).}

Assuming that bound-state $\beta$-decay of tritium to the ground state of $\HeThree$
exists and occurs at the half-a-percent level, the reverse process can occur resonantly
if no energy is lost to nuclear recoil in either the emission or absorption of the
$\bar{\nu}_e$.  The ME achieves this by embedding the nuclei in a crystal lattice,
and allowing the entire macroscopic lattice to absorb the recoil momentum (but no recoil energy).
While simple to say in words, the actual physical implementation of such an effect
is difficult due to the conflicting requirements for the ME to occur.
On the one hand, both the emitter ($\tritium$) and absorber ($\HeThree$) nuclei
must be bound and fixed to the crystal lattice for recoilless emission/absorption to
occur.  
On the other hand, the same interactions responsible for this tend to
broaden the width of the decay and attenuate the resonance.
Doppler effects due to temperature can also lead to width broadening, 
and gravitational potentials will detune the resonance.\footnote{%
The ME was how gravitational red shift was first measured, after all.}
The different chemical properties of $\tritium$ and $\HeThree$, the latter
being a noble gas, also makes it difficult to achieve the exact same environment
for the emitter and absorber: $\tritium$ forms chemical bonds while $\HeThree$ does not.

In a series of papers starting in 2005 \cite{Raghavan:2005gn,Raghavan:2006xf,Raghavan:2009hj,Raghavan:2009dk,Raghavan:2009zz,Raghavan:2009sf},
Raghavan argued that the ME could be achieved by taking advantage of the
fact that tritium forms interstitial compounds, aka ``tritides,'' with metals, 
in which the tritium atoms are trapped inside the grid of the metallic crystal \cite{Lasser:1989}.\footnote{%
The study of embedding hydrogen atoms inside metallic crystals has been pursued
as a technology for hydrogen storage.}
Raghavan proposed Niobium tritide as the most promising candidate. 
By allowing the tritium in Niobium tritide to $\beta$-decay, one could replace them with
$\HeThree$ which would be distributed in the same environment in the same way as the tritium.
He argued that due to the very long lifetime of tritium \cite{Lucas:2000}, 
the effects of lattice vibrations would cancel out and not lead to
width broadening, and result in the tritium 
decay being detected with its very narrow natural width of $\Gamma\sim 10^{-24}\,\mathrm{eV}$ \cite{Raghavan:2009hj,Raghavan:2009dk,Raghavan:2009zz,Raghavan:2009sf}
and the cross section enhanced to $\sigma\sim 10^{-17}\,\mathrm{cm}^2$ on resonance.

Raghavan's assessment of the effect of lattice vibrations had been contested
by several groups who argued that the lattice vibrations do indeed lead to
width broadening and an attenuation of the resonance strength \cite{Potzel:2009qe,Potzel:2009pr,Potzel:2011zza,Schiffer:2009zz}.  
Raghavan was still quite confident of his result, however, as you can judge from his
tone in section 3.
Another objection came from Suzuki et al. \cite{Suzuki:2010zza} who pointed
out that with the tritium densely populating the Niobium lattice, its $1s$ 
orbitals would form a band structure, again broadening the resonance width.

Due to these possible difficulties, 
Potzel \cite{Potzel:2011zza} has recently argued for the following
Holmium163 ($\Ho$) $\leftrightarrow$ Dysprosium163 ($\Dy$) pair
as an alternative M\"ossbauer neutrino system:
\begin{equation}
\Ho\,(4f^{11}) \quad\leftrightarrow\quad \Dy\,(4f^{10}) + \nu_e\;,\qquad (Q=2.6\,\mathrm{keV})\;.
\end{equation}
This system was also listed by Kells and Schiffer \cite{Kells:1984nm} as a possible
candidate.
Atomic $\Dy$ is stable.  However, when stripped of all of its
electrons it can $\beta$-decay into $\Ho$.
Curiously, this was the first reaction in which bound-state $\beta$-decay,
\begin{equation}
\Dy^{66+} \quad\rightarrow\quad \Ho^{66+} + \bar{\nu}_e\;,
\end{equation}
was experimentally confirmed
at the GSI in 1992 \cite{Jung:1992pw}.

\subsection{The Time-Energy Uncertainty Relation}

Uncertainty relations play an important role in the quantum theory. They
are based on fundamental general properties of the theory and manifest the
nature of it. There are two different types of the uncertainty relations in
the quantum theory: Heisenberg uncertainty relations and time-energy uncertainty
relations.

The Heisenberg uncertainty relations are based on commutation relations
for hermitian operators, which correspond to physical quantities. 
Let us consider two hermitian operators $\hat{A}$ and $\hat{B}$. 
From the Cauchy-Schwarz inequality we have
\begin{equation}
(\Delta A)_\psi\;(\Delta B)_\psi \;\ge\; \dfrac{1}{2}
\left|\bra{\psi} \,[\hat{A},\hat{B}]\, \ket{\psi} \right|
\;,
\label{ABuncert}
\end{equation}
where $\ket{\psi}$ is some state, and
\begin{equation}
(\Delta A)_\psi \;=\; 
\sqrt{\bra{\psi}\hat{A}^2\ket{\psi} - \bra{\psi}\hat{A}\ket{\psi}^2}
\end{equation}
is the standard deviation of the probability distribution of
possible outcomes of a measurement of $\hat{A}$ on the state $\ket{\psi}$.\footnote{%
Note that the `uncertainty' $(\Delta A)_\psi$ 
in the measurement of $\hat{A}$ on state $\ket{\psi}$
has nothing to do with the `disturbance' of the state $\ket{\psi}$
by the measurement process itself as originally formulated by Heisenberg
in his famous Heisenberg microscope thought experiment \cite{Heisenberg:1927}.
See Ref.~\cite{Ozawa:2003} for a clarification of this distinction.}
For the canonical pair $\hat{x}$ and $\hat{p}$, 
we have $[\hat{x},\hat{p}]=i\hbar$, and
the lower bound will be independent of the state $\ket{\psi}$:
\begin{equation}
\Delta x\,\Delta p \;\ge\; \dfrac{\hbar}{2}\;.
\label{xpuncert}
\end{equation}
The time-energy uncertainty relation 
\begin{equation}
\Delta t\,\Delta E \;\ge\; \dfrac{\hbar}{2}
\label{tEuncert}
\end{equation}
has a completely different character. 
While deduced immediately from Eq.~(\ref{xpuncert})
by demanding relativistic covariance, it does not share the same operator-based
foundation, the time $t$ being a parameter rather than an operator.
Consequently, it has been a subject of intense discussions and controversy from the early years of the quantum theory. 
In the literature exist many different derivations of Eq.~(\ref{tEuncert}), each 
with different meanings of the quantities that enter into them \cite{Mandelstam:1945,Aharonov:1961,Aharonov:1975,Anandan:1990fq,busch:1990aa,busch:1990bb,busch:1990cc}.

For instance, Mandelstam and Tamm \cite{Mandelstam:1945} start from the
Heisenberg equation of motion
\begin{equation}
i\hbar\dfrac{d}{dt}\,\hat{A}(t)\;=\; [\,\hat{A}(t),\,\hat{H}\,]\;,
\end{equation}
to which an application of Eq.~(\ref{ABuncert}) yields
\begin{equation}
(\Delta A)_\psi(\Delta E)_\psi \;\ge\;
\dfrac{\hbar}{2}
\left|\dfrac{d}{dt}\bra{\psi}\hat{A}(t)\ket{\psi}
\right|
\;.
\end{equation}
Defining
\begin{equation}
(\Delta t)_\psi \;\equiv\;
\dfrac{(\Delta A)_\psi}{
\left|\dfrac{d}{dt}\bra{\psi}\hat{A}(t)\ket{\psi}
\right|
}\;,
\end{equation}
they obtain $(\Delta t)_\psi (\Delta E)_\psi \ge \hbar/2$
with a very particular meaning of the uncertainty $(\Delta t)_\psi$:
it is the time interval which is necessary for the
expectation value $\bra{\psi}\hat{A}(t)\ket{\psi}$ to be changed by one standard deviation $(\Delta A)_\psi$.
In other words $(\Delta t)_\psi$ characterizes the time interval during which the state of
the system significantly varies.

The most straightforward derivation starts from the Schr\"odinger equation
itself \cite{Anandan:1990fq}
\begin{equation}
i\hbar\dfrac{d}{dt}\,\psi(t) \;=\; \hat{H}\psi(t)\;,
\end{equation}
which associates the eigenvalues of the Hamiltonian $\hat{H}$ with the
oscillational frequency of the wave-function $\psi(t)$.
The relation $\Delta t\,\Delta E \ge \hbar/2$ is then 
the standard relation between the spread in time and the spread in frequency
for any wave phenomenon.
Indeed, in the context of quantum field theory (QFT), in which the position $x$ is also a parameter like $t$ and not an operator, the relation $\Delta x\,\Delta p\ge \hbar/2$ is 
also the standard relation between the spread in space and the spread in wavenumber
for any wave phenomenon,
and both uncertainty relations are placed on an equal footing.

The uncertainty in energy of an eigenstate of $\hat{H}$ created at $t=0$ 
would not be zero at any finite time $t=\Delta t$ since no wave of finite duration
can have a definite frequency.  Rather, its frequency will have a spread of
$\Delta E\sim \hbar/\Delta t$ which can only go to zero in the limit 
$t=\Delta t\rightarrow \infty$.
Thus, the `uncertainty' $\Delta t$ should be interpreted as the duration of
the state since its creation.  
If the state is unstable with lifetime $\tau$, 
then as $t\rightarrow\infty$ the uncertainty in the frequency can be
expected to settle to its `natural' value of $\Delta E_\mathrm{nat}\sim \hbar/\tau$.
This was the phenomenon observed by Wu et al. \cite{wu:1960},
Lynch et al. \cite{lynch:1960} and others utilizing the ME in $\gamma$-rays,
and which Raghavan proposed to observe using the ME in neutrinos.

\section{Experimental Test of the Time-Energy Uncertainty Relation with\\
M\"ossbauer Neutrinos -- by R. S. Raghavan}

One of the most basic predictions of quantum mechanics (QM) is the time-energy uncertainty (TEU) relation 
$\Delta E\,\Delta t \ge \hbar/2$, in parallel to the Heisenberg relation 
$\Delta p\,\Delta x \ge \hbar/2$.  
Since time is not a canonical variable like $\hat{p}$ and $\hat{x}$ for writing out a commutator that underlies the uncertainty product, the meaning of the TEU has been debated for decades, even though it has been verified in numerous atomic, nuclear and particle decays -- the contexts for the role of the TEU which sets the minimum observable energy width $\Delta E = \hbar/\tau$.
The energy-time relation has been tested in numerous experiments \cite{auletta}. In this note we suggest a neutrino analog of the famous test of the
energy-time relation performed in \cite{wu:1960}.

It is not clear however, whether the TEU still holds in the limit of extremely small widths of very long lived states where other factors may enter in altering the minimum observable energy.  For example, Mead \cite{Mead:1966zz} has argued that the ultimate measurable energy width could be limited by the fundamental (Planck) length $\ell_P = \sqrt{G\hbar/c^3}$ ($G=\mbox{Newton's constant}$) ahead of that expected from the TEU (via an independently measured state lifetime). He suggests that such an effect -- a Planck broadening of spectral lines -- may be observable with energy precision $\Delta E/E\sim 10^{-20}$ to $10^{-40}$.
If there is an experimental tool to probe this extreme regime (recent work suggests the possibility), the question arises: If it is found that the Planck limited width is broader than the QM width, what is the role of time (i.e. the measured lifetime) and the TEU itself which would be in conflict with the result? This is the question addressed in this paper. We discuss its deep ramifications. We offer a generalization of the TEU as 
$\Delta E\,\Delta t \ge (\hbar/2)[1+(\Delta t/\Tstar)^n]$ where $\Delta t$ is the time of measurement (in this case the lifetime of the state $\tau$, in the present case $\sim 10^9\,\mathrm{s}$), $\Tstar\sim \Lstar/c$ the time for light to cross the Universe $\sim 3\times 10^{18}\,\mathrm{s}$,  
and $n$ is a parameter subject to adjustment with future measurements.\footnote{%
Raghavan's idea was that if the large time scale $\Tstar$ has its origins in quantum gravity,
then it would be related to the size of the observable universe $\Lstar$.
The size of the observable universe is currently $\Lstar\sim 93\mbox{ billion light years}\sim 10^{27}\,\mathrm{m}$, which yields
$\Tstar = \Lstar/c \sim 3\times 10^{18}\,\mathrm{s}$.
}
Note that this generalization preserves the traditional QM foundations\footnote{%
As we will argue in section~4, it will actually rock the very foundations of QM.
}
but introduces parameters that can be connected in the future to the Planck length and/or quantum gravity.

Recently, however, a precision of $\Delta E/E\sim 5\times 10^{-29}$ appears possible by the newfound neutrino resonance reaction $\tritium\leftrightarrow \HeThree$ (tritium $\tritium$: lifetime $\tau\sim 18\,\mathrm{yrs}$ \cite{Lucas:2000}, TEU width $\Gamma(\mathrm{TEU})\sim \hbar/\tau \sim 10^{-24}\,\mathrm{eV}$), which offers a meaningful TEU test in an extreme energy-time regime not probed up to now. The purpose of this note is to discuss first, a direct test of TEU in tritium decay.  Looking ahead to the possibility of a deviation from the quantum theoretical TEU as conjectured by Mead, we explore generalization of the TEU relation on foundations beyond quantum theory and their connections to the fundamental length $\ell_P$ and quantum gravity. 
While generalized forms of $\Delta p\,\Delta x \ge \hbar/2$ have been discussed in the context of quantum gravity (see \cite{Maggiore:1993kv,Kempf:1994su,Kempf:1996fz,Chang:2001kn,Chang:2001bm,Benczik:2002tt,Benczik:2005bh,Chang:2010ir,Chang:2011jj,Lewis:2011fg} and references therein), 
we are aware of fewer such attempts \cite{Shalyt-Margolin:2002} regarding the TEU.

The $\bar{\nu}_e$ resonance reaction in the case of 2-body $\beta$-decay of $\tritium$, which results in emission of a mono-energetic 18.6 keV neutrino ($\bar{\nu}_e$), is reminiscent of the well-known $\gamma$-ray resonance [M\"ossbauer Effect (ME)]. But the new aspect is the long lifetime of $\tritium$ which presents entirely new facets well beyond those of classical ME \cite{Raghavan:2009hj}. 
The significant and surprising discovery \cite{Raghavan:2009hj} is that, in this case, 
a useful fraction of the $\bar{\nu}_e$ will always be emitted with the \textit{natural width}, 
$\hbar/\tau\sim 10^{-24}\,\mathrm{eV}$ in spite of myriad extra-nuclear sources of broadening and energy shifts well known in the ME work \cite{Potzel:2009qe}. 
The reason is that for long lifetimes the fluctuations act to modulate and average out the fluctuations rather than detune the $\bar{\nu}_e$ energy. 
Thus, they affect only the intensity of the resonance signal, not the width as they do in short lived ME levels \cite{Raghavan:2009sf}.\footnote{%
\textbf{Footnote by Raghavan:}
B.~Balko et al. (Priv comm. and to be published) have unequivocally confirmed theoretically the claims of Raghavan in ref.~\cite{Raghavan:2009hj} and \cite{Raghavan:2009sf}.
} 
The maximum resonance cross section $\sigma_0$ (height of the resonance line) is governed by the flux density of $\bar{\nu}_e$ emitted within the resonance width $\Delta E$. 
In the case of the natural width $\Delta E_{\mathrm{nat}}$, $\sigma_0$ 
takes the maximum value of $\sigma_0=2\pi\lambdabar$ \cite{Potzel:2009qe} where $\lambdabar$ is the $\bar{\nu}_e$ wavelength. 
In the ${}^3\mathrm{H}\leftrightarrow \HeThree$ reactions, $\sigma_0$ is very large, $\sim 10^{-17}\,\mathrm{cm}^2$.  
The $\bar{\nu}_e$'s are resonantly absorbed in $\HeThree$,  
back transforming it to $\tritium$ which then $\beta$-decays. 
The simplest resonance signal is electron emission of the resonantly induced activity.  The percent absorption at zero detuning of the resonance, i.e. the rate of resonant activation is the direct experimental handle on the line width.  The energy sensitivity of the resonance is $\Delta E/E = 10^{-24}/18.6\times 10^{4} = 5.4\times 10^{-29}$.

The basic questions in the application of the TEU relation $\Delta E\,\Delta t \ge \hbar/2$ is the definition of the ``time'' $\Delta t$ in the TEU, the resulting energy width $\Delta E$ and the effect on $\sigma_0$.\footnote{%
\textbf{Footnote by Raghavan:}
Surprisingly confusing opinions on the definition of ``time'' have been expressed. 
Lipkin \cite{Lipkin:2009uq} asserts that the TEU with his definition of the ``time'' as the time of flight of the $\bar{\nu}_e$ would make it impossible to observe the $\bar{\nu}_e$ resonance. 
This claim contradicts numerous ME experiments performed with different baselines with no effect on the resonance. 
Potzel and Wagner \cite{Potzel:2009qe} assert that the TEU does not play any role in the $\bar{\nu}_e$ resonance line width, 
while Akhmedov, Kopp, and Lindner \cite{Akhmedov:2008jn} conclude that an imprecisely defined ``time of the experiment'' does. 
Both claims are refuted by early ME experiments that address this specific question (see text).}
The roles of these observables have been particularly clarified by the ME experiments by Wu et al. and others \cite{wu:1960,lynch:1960}, and by theory \cite{harris:1960}.
The experiments used the classic decay $\Co\rightarrow \Fe$ 
to an initial state at 136~keV in the daughter which decays unmeasurably fast to the isomer which emits the well known 14.4~keV ME $\gamma$-ray with a mean life of $\tau\sim 144\,\mathrm{ns}$. 
The experiment detected both the 122~keV precursor and the 14.4~keV photons.  
The ME of the latter was observed in coincidence with the 122~keV photon. 
The 122~keV signal marks the creation of the 14.4~keV state and a known delay $T$ between the two photon signals measures the duration for which the 14.4~keV state remains unperturbed. 
The detection of the 14.4~keV photon marks the end of the time of measurement. 
The resonance signal detected after this time $T$ measured the precise energy of photon (and the emitting state) with an uncertainty given by the width of the resonance. 
The results showed that the width depended on $T$.  
At short delays $T$, the width was broad and as $T$ approached the lifetime $\tau$ of the state the width narrowed to the natural line-width of the state. The absorption at zero detuning quantitatively measures the resonance width by proxy. Complete agreement with TEU was found in these remarkably clear results. 
This experiment also clearly illustrates the definitions of $T$ and $E$ in the TEU. 
To repeat, the ``time'' in the TEU is the unperturbed duration of the unstable state starting from its creation and lasting till the state decays, as detected by its decay product. 

\begin{figure}[t]
\begin{center}
\includegraphics[width=10cm]{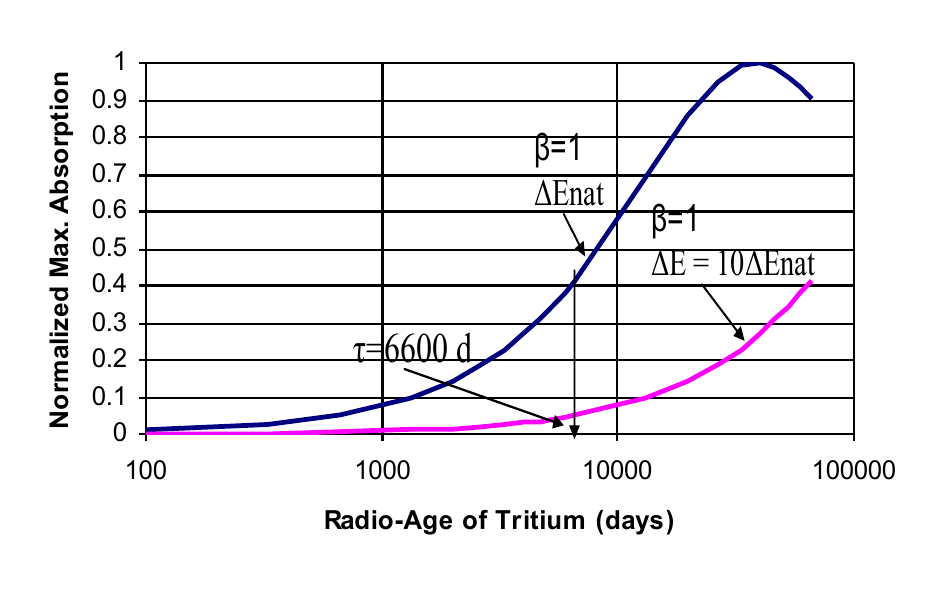}
\end{center}
\caption{Growth rates of normalized resonance absorption $A$ with radio-age of $\tritium$ for QM-TEU.  
Left curve for $\Delta E(\mbox{QM})=\Delta E_{\mathrm{nat}}$, for the measured $\tau$ of $\tritium$ for  known absorber thickness $\beta=1$.  Right curve for hypothetical $\Delta E(\mbox{non-QM})=10\,\Delta E_{\mathrm{nat}}$ for $\beta=1$.}
\label{rajufig1}
\end{figure}

This ``time-filtered'' resonance method can be applied directly without electronic coincidences to the the 
$\bar{\nu}_e$ resonance of $\tritium\leftrightarrow\HeThree$ because of the long lifetime of the 
$\tritium$ state. 
The time starts with the creation of tritium (in a reactor or elsewhere in a time short compared to the lifetime of $\tritium$) and ends with the detection of the resonance signal $\HeThree\rightarrow\tritium$.
It does not start with installation of the $\tritium$ in the experimental set up and starting the counting.  
It is important to realize that in the TEU framework it is immaterial whether the created $\tritium$ is stored for a long time before installation in the experiment. 
The total time after creation (storage+installation) up to the detection of the neutrino is the delay time $T$, thus the resonance signal strength depends on the age of the tritium, a novel effect demanded by the TEU. 
By the same token, as the counting proceeds, the delay time increases naturally, the effective resonance line-width decreases via the TEU, thus the resonance signal spontaneously increases with time. 
Observation of these time dependences is the most straightforward experimental demonstration of the TEU (see Fig.~\ref{rajufig1}).

\section{Theoretical Implications - Minic, Takeuchi, Tze}

The above discussion was concerned with a new fundamental neutrino probe of the
energy-time relation in quantum theory. Now we would like to discuss 
possible theoretical implications of such a new probe of the foundations
of quantum theory.
Given that the proposed experiment is sensitive to small deviations
from the standard energy-time uncertainty relation
\begin{equation}
\Delta t\,\Delta E \;\ge\; \hbar/2\;,  
\label{ETUR}
\end{equation}
one must ask in what way the physics at the Planck scale $\ell_P$
could modify this relation.


First and foremost,
in the presence of a fundamental length scale $\ell_P$ 
one expects spatial resolutions to be bounded from below 
$\Delta x \ge \ell_P$ \cite{Padmanabhan:1987au}.
It has been argued by Mead \cite{Mead:1966zz} that 
this finiteness in $\Delta x$ will naturally lead to an
uncertainty in the energy of a particle trapped in a potential $V$ of order
\begin{equation}
\Delta E 
\;\sim\; \Delta\mathbf{x}\cdot(\bm{\nabla}V)  
\;\agt\; \ell_P\left(\dfrac{\partial V}{\partial r}\right)\;,
\end{equation}
in addition to any intrinsic energy uncertainty the state may have.
If $R$ denotes the characteristic size of a bound system (such as
the size of the nucleus), and $\nu_0$ the average frequency of the 
characteristic transition, then one expects a frequency shift of
$
\Delta \nu \ge (\ell_P/R)\,\nu_0.
$
Unfortunately, as long as one works within the confines of 
canonical quantum theory, one cannot disentangle
the uncertainty due to $\ell_P$ from the intrinsic one.
In particular, the energy-time relation, Eq.~(\ref{ETUR}), 
being a consequence of the Schr\"{o}dinger equation, remains unaffected.

Attempts to incorporate the effect of $\ell_P$ into quantum theory
has been through the deformation of the canonical commutation relation
between $\hat{x}$ and $\hat{p}$ to \cite{Maggiore:1993kv,Kempf:1994su,Kempf:1996fz}
\begin{equation}
[\,\hat{x},\,\hat{p}\,] \;=\; i\hbar\left( 1 + \alpha\, \hat{x}^2 + \beta\, \hat{p}^2 + \cdots\right)\;,
\end{equation}
where the ellipses denote possible higher order terms in $\hat{x}$ and $\hat{p}$ which are neglected.
This leads to the relation
\begin{equation}
\Delta x\;\Delta p \;\ge\;\dfrac{\hbar}{2}
\left[\, 1 + \alpha (\Delta x)^2 + \beta (\Delta p)^2 \,\right]\;,
\label{DeformedUR}
\end{equation}
assuming $\alpha\langle \hat{x}^2 \rangle + \beta\langle \hat{p}^2 \rangle = 0$.
On dimensional grounds one expects $\beta\sim \ell_P^2/\hbar^2$,
while $\alpha\sim \Lstar^{-2}$ where $\Lstar$ must be a very large length scale
(the size of the universe?) to prevent obvious phenomenological problems.  
The appearance of a large scale $\Lstar$ from Planck-scale physics
may be justifiable through the UV/IR mixing nature expected of
quantum gravity \cite{Chang:2010ir}.
The $\alpha=0$ case is motivated by perturbative string theory \cite{Gross:1987ar,Gross:1987kza,Amati:1988tn,Amati:1987uf,Amati:1987wq,Witten:1997fz}
and leads to the lower bound $\Delta x \ge \hbar\sqrt{\beta} \sim \ell_P$.
This deformation of $[\,\hat{x},\,\hat{p}\,]$ can 
maintain consistency within canonical non-relativistic quantum theory,
and thereby also maintain Eq.~(\ref{ETUR}),
and various phenomenological consequences have been worked out \cite{Chang:2001kn,Chang:2001bm,Benczik:2002tt,Benczik:2005bh,Chang:2010ir,Chang:2011jj,Lewis:2011fg}.
However, if we insist on rendering Eq.~(\ref{DeformedUR}) covariant,
Eq.~(\ref{ETUR}) must also be modified to
\begin{equation}
\Delta t\; \Delta E \;\ge\; 
\dfrac{\hbar}{2}\left[\, 1+ \alpha (\Delta t)^2 + \beta (\Delta E)^2 \,\right]\;,
\label{ETUR-deformed}
\end{equation}
which would require a (non-local) modification of the Schr\"{o}dinger equation \cite{camelia}.
Such a generalization of the Schr\"{o}dinger equation may
occur in a more fundamental approach to quantum gravity \cite{Minic:2003en,Minic:2003nx,Minic:2004rj,Jejjala:2006jf,Jejjala:2007rn,Jejjala:2008nc,Jejjala:2009ky}, but it is
conceptually and technically a very challenging task.
As mentioned above, we expect the parameter $\alpha$ to correspond to the
inverse square of some very long time scale.
If we accept Eq.~(\ref{ETUR-deformed}) purely on phenomenological grounds,  
$\alpha$ could be observed, or bounded, in the experiment proposed in this paper, in particular, 
the neutrino analog of the Wu et al. experiment \cite{wu:1960}.


Given the rigidity of canonical quantum theory, 
a relation such as Eq.~(\ref{ETUR-deformed}) which would require a major deformation of the
Schr\"odinger equation itself may be too radical to contemplate.
So let us consider a more conservative option.
Suppose that the Hamiltonian of some more fundamental theory,
such as a quantum theory of gravity, depends on some large fundamental time-scale, $\Tstar$, 
and that the energy uncertainty is shifted from $\Delta E$ to $\Delta E - (\hbar/\Tstar)$
so that the corrected energy-time uncertainty relation reads
\begin{equation}
\Delta t\; \Delta E \;\sim\; \hbar\left( 1 + \frac{\Delta t}{\Tstar}\right)\;.
\end{equation}
The sign of the shift in $\Delta E$ has been chosen to maintain the positivity of the right-hand side.
This corrected energy-time relation would be reasonable from the point of view of the proposed experiment for large $\Delta t$. Such a linear shift in time can be accomplished while preserving unitary evolution by way of a suitable 
$\Tstar$-dependent operator.


Finally, another possibility is  the ``inverse'' of the preceding modification of the energy time relation
\begin{equation}
\Delta t\; \Delta E \;\sim\; \hbar\left(1 + \dfrac{\taustar}{\Delta t}\right)\;,
\label{eq7}
\end{equation}
where $\taustar$ must be small to prevent phenomenological problems at small $\Delta t$.
Such an expression can be readily obtained from the following local linear generalization of the Schr\"{o}dinger equation
\begin{equation}
\hat{H} \psi \;=\; i\hbar\,(\delta_t \psi) - i \hbar\, \taustar (\delta_t^2 \psi)\;.
\end{equation}
Being diffusive, such a modified Hamiltonian is non-unitary.
Moreover, the change from the standard energy-time relation, Eq.~(\ref{ETUR}),
would unfortunately \textit{not} be observable in the above proposed experiment, as the fundamental scale $\taustar$ should be very short.

In the context of the preceeding modification of the Schr\"{o}dinger equation, it is tempting to speculate that some new quantum gravitational degrees of freedom \cite{Minic:2003en,Minic:2003nx,Minic:2004rj,Jejjala:2006jf,Jejjala:2007rn,Jejjala:2008nc,Jejjala:2009ky} are responsible for the overall
conservation of probability, and that the proper inclusion of
these degrees of freedom would restore a seemingly violated unitary evolution.
If indeed quantum gravity is essential for the problem
of quantum measurement \cite{penrose}, then the last model, Eq.~(\ref{eq7}), for the generalized
time-energy uncertainty relation presented in this note
would be a manifestation of such fundamentally new physics.

\section{Closing Remarks}

Raju Raghavan was a remarkably creative physicist. 
His novel suggestion concerning the M\"ossbauer neutrinos 
can be used to make neutrinos the next natural probe of the foundations of quantum theory.
In this note, we have proposed how neutrinos could be used to probe the 
time-energy uncertainty relation which is ultimately related to 
unitary time evolution of quantum states. 
Any departures from this fundamental relation would be indicative
of a new fundamental framework for quantum physics, which may be required 
by the quantum theory of gravity \cite{Chang:2011yt}.
Thus, neutrinos can be seen as the next natural probe not only of the early history of
the Universe, via the cosmic neutrino background \cite{weinberg:cosmo}, but also of the 
fundamental quantum underpinnings of the Universe itself.
We sincerely hope that future avatars of Raghavan 
will further develop this fascinating research direction.

\acknowledgments

We wish to thank Patrick Huber and Jonathan Link for useful discussions 
about the physics of this proposal. 
Special thanks go to Bruce Vogelaar for his invaluable help and encouragement. 
The collaboration of Bohdan Balko with Raju Raghavan is also gratefully acknowledged.
This work was supported in part by the U.S. Department of Energy, grant number DE-FG05-92ER40677, task A (DM and TT) and by
the World Premier International Research Center Initiative, MEXT, Japan (TT).



\begin{thebibliography}{99}


\bibitem{Raghavan:2005gn} 
  R.~S.~Raghavan,
  ``Recoilless resonant capture of antineutrinos,''
  hep-ph/0511191.

\bibitem{Raghavan:2006xf} 
  R.~S.~Raghavan,
  ``Recoilless resonant capture of antineutrinos from tritium decay,''
  hep-ph/0601079.

\bibitem{Raghavan:2009hj} 
  R.~S.~Raghavan,
  ``Hypersharp Resonant Capture of Neutrinos as a Laboratory Probe of the Planck Length,''
  Phys.\ Rev.\ Lett.\  {\bf 102}, 091804 (2009)
  [arXiv:0903.0787 [hep-ph]];

\bibitem{Raghavan:2009dk} 
  R.~S.~Raghavan,
  ``Time-Energy Uncertainty in Neutrino Resonance: Quest for the Limit of Validity of Quantum Mechanics,''
  arXiv:0907.0878 [hep-ph].

\bibitem{Raghavan:2009zz} 
  R.~S.~Raghavan,
  ``Reply to 'Comment on 'Hypersharp Resonant Capture of Neutrinos as a Laboratory Probe of the Planck Length'',''
  Phys.\ Rev.\ Lett.\  {\bf 103}, 099103 (2009).

\bibitem{Raghavan:2009sf} 
  R.~S.~Raghavan,
  ``Why Neutrino Lines are Hypersharp,''
  arXiv:0908.2980 [hep-ph].


\bibitem{wu:1960}
  C.~S.~Wu, Y.~K.~Lee, N.~Benczer-Koller, and P.~Simms,
  ``Frequency Distribution of Resonance Line Versus Delay Time,''
  Phys.\ Rev.\ Lett.\ {\bf 5}, 432 (1960).
  
\bibitem{lynch:1960}
  F.~J.~Lynch, R.~E.~Holland, and M.~Hamermesh,
  ``Time Dependence of Resonantly Filtered Gamma Rays from Fe${}^{57}$,''
  Phys.\ Rev.\ {\bf 120}, 513 (1960).

\bibitem{harris:1960}
  S.~M.~Harris,
  ``Quantum Mechanical Calculation of M\"ossbauer Transmission,'
  Phys.\ Rev.\ {\bf 124}, 1178 (1960).


\bibitem{Mossbauer:1958aa}
  R.~L.~M\"ossbauer,
  ``Kernresonanzfluoreszenz von Gammastrahlung in Ir191,''
  Z.\ Phys.\ {\bf 151}, 124 (1958).

\bibitem{Mossbauer:1958bb}
  R.~L.~M\"ossbauer,
  ``Kernresonanzabsorption von Gammastrahlung in Ir191,''
  Naturwissenschaften {\bf 45}, 538 (1958).
  
\bibitem{Mossbauer:1959}
  R.~L.~M\"ossbauer,
  ``Kernresonanzabsorption von $\gamma$-Strahlung in Ir191,''
  Z.\ Naturforschung\ {\bf 14a}, 211 (1959).

\bibitem{Mossbauer:1962xd} 
  R.~L.~M\"ossbauer,
  ``Recoilless nuclear resonance absorption,''
  Ann.\ Rev.\ Nucl.\ Part.\ Sci.\  {\bf 12}, 123 (1962).




\bibitem{Mead:1964zz} 
  C.~A.~Mead,
  ``Possible Connection Between Gravitation and Fundamental Length,''
  Phys.\ Rev.\  {\bf 135}, B849 (1964).

\bibitem{Mead:1966zz}
  C.~A.~Mead,
  ``Observable Consequences of Fundamental-Length Hypotheses,''
  Phys.\ Rev.\  {\bf 143}, 990 (1966).


\bibitem{Andryushin:1973qk} 
  V.~I.~Andryushin and V.~N.~Melnikov,
  ``M\"ossbauer effect on neutrino and limits on elementary length,''
  Lett.\ Nuovo Cim.\  {\bf 7}, 809 (1973).


\bibitem{Kells:1981xt} 
  W.~P.~Kells,
  ``Can recoilless nuclear anti-neutrino emission be usefully detected?,''
  FERMILAB-FN-0340 (1981).

\bibitem{Kells:1982rm} 
  W.~P.~Kells,
  ``Resonant neutrino activation and neutrino oscillations,''
  AIP Conf.\ Proc.\  {\bf 99}, 272 (1983).

\bibitem{Kells:1984nm} 
  W.~P.~Kells and J.~P.~Schiffer,
  ``Possibility of Observing Recoilless Resonant Neutrino Absorption,''
  Phys.\ Rev.\ C {\bf 28}, 2162 (1983).

\bibitem{Schiffer:2009zz} 
  J.~P.~Schiffer,
  ``Comment on 'Hypersharp Resonant Capture of Neutrinos as a Laboratory Probe of the Planck Length',''
  Phys.\ Rev.\ Lett.\  {\bf 103}, 099102 (2009).


\bibitem{Potzel:2006ad} 
  W.~Potzel,
  ``Recoilless resonant capture of antineutrinos: Basic questions and some ideas,''
  Phys.\ Scripta T {\bf 127}, 85 (2006).

  
\bibitem{Potzel:2009qe} 
  W.~Potzel and F.~E.~Wagner,
  ``Comment on 'Hypersharp Resonant Capture of Neutrinos as a Laboratory Probe of the Planck Length',''
  Phys.\ Rev.\ Lett.\  {\bf 103}, 099101 (2009),
  [arXiv:0908.3985 [hep-ph]].

\bibitem{Potzel:2009pr} 
  W.~Potzel,
  ``M\"ossbauer antineutrinos: Some basic considerations,''
  Acta Phys.\ Polon.\ B {\bf 40}, 3033 (2009)
  [arXiv:0912.2221 [hep-ph]].

\bibitem{Potzel:2011zza} 
  W.~Potzel,
  ``M\"ossbauer Antineutrinos: Recoilless Resonant Emission and Absorption of Electron Antineutrinos,''
  Phys.\ Part.\ Nucl.\  {\bf 42}, 661 (2011)
  [arXiv:1012.5000 [hep-ph]].


\bibitem{Suzuki:2010zza} 
  D.~Suzuki, T.~Sumikama, M.~Ogura, W.~Mittig, A.~Shiraki, Y.~Ichikawa, H.~Kimura, H.~Otsu, H.~Sakurai, Y.~Nakai and M.~S.~Hussein,
  ``Resonant neutrino scattering: An impossible experiment?,''
  Phys.\ Lett.\ B {\bf 687}, 144 (2010).



\bibitem{Lipkin:2009uq} 
  H.~J.~Lipkin,
  ``Difficulties in using the sharp neutrino spectrum at short times,''
  arXiv:0904.4913 [hep-ph].


\bibitem{Bilenky:2008ez} 
  S.~M.~Bilenky, F.~von Feilitzsch and W.~Potzel,\\
  ``Time-Energy Uncertainty Relations for Neutrino Oscillation and M\"ossbauer Neutrino Experiment,''
  J.\ Phys.\ G {\bf 35}, 095003 (2008)
  [arXiv:0803.0527 [hep-ph]].

\bibitem{Bilenky:2008dk} 
  S.~M.~Bilenky, F.~von Feilitzsch and W.~Potzel,
  ``Different Schemes of Neutrino Oscillations in M\"ossbauer Neutrino Experiment,''
  arXiv:0804.3409 [hep-ph].
  
\bibitem{Bilenky:2009zz} 
  S.~M.~Bilenky, F.~von Feilitzsch and W.~Potzel,
  ``Reply to the Comment on `On application of the time-energy uncertainty relation to M\"ossbauer neutrino experiments' by E.~Kh.~Akhmedov, J.~Kopp and M.~Lindner,''
  J.\ Phys.\ G {\bf 36}, 078002 (2009).

\bibitem{Bilenky:2011pk} 
  S.~M.~Bilenky, F.~von Feilitzsch and W.~Potzel,
  ``Neutrino Oscillations and Uncertainty Relations,''
  J.\ Phys.\ G {\bf 38}, 115002 (2011)
  [arXiv:1102.2770 [hep-ph]].


\bibitem{Akhmedov:2008jn} 
  E.~Kh.~Akhmedov, J.~Kopp and M.~Lindner,
  ``Oscillations of M\"ossbauer neutrinos,''
  JHEP {\bf 0805}, 005 (2008)
  [arXiv:0802.2513 [hep-ph]].

\bibitem{Akhmedov:2008zz} 
  E.~Kh.~Akhmedov, J.~Kopp and M.~Lindner,
  ``On application of the time-energy uncertainty relation to M\"ossbauer neutrino experiments,''
  J.\ Phys.\ G {\bf 36}, 078001 (2009)
  [arXiv:0803.1424 [hep-ph]].

\bibitem{Kopp:2009fa} 
  J.~Kopp,
  ``M\"ossbauer neutrinos in quantum mechanics and quantum field theory,''
  JHEP {\bf 0906}, 049 (2009)
  [arXiv:0904.4346 [hep-ph]].

\bibitem{Minakata:2006ne} 
  H.~Minakata and S.~Uchinami,
  ``Recoilless resonant absorption of monochromatic neutrino beam for measuring Delta m**2(31) and theta(13),''
  New J.\ Phys.\  {\bf 8}, 143 (2006)
  [hep-ph/0602046].

\bibitem{Minakata:2007tn} 
  H.~Minakata, H.~Nunokawa, S.~J.~Parke and R.~Zukanovich Funchal,
  ``Determination of the neutrino mass hierarchy via the phase of the disappearance oscillation probability with a monochromatic anti-electron-neutrino source,''
  Phys.\ Rev.\ D {\bf 76}, 053004 (2007)
  [Erratum-ibid.\ D {\bf 76}, 079901 (2007)]
  [hep-ph/0701151].


\bibitem{Parke:2008cz} 
  S.~J.~Parke, H.~Minakata, H.~Nunokawa and R.~Z.~Funchal,
  ``Mass Hierarchy via Mossbauer and Reactor Neutrinos,''
  Nucl.\ Phys.\ Proc.\ Suppl.\  {\bf 188}, 115 (2009)
  [arXiv:0812.1879 [hep-ph]].

\bibitem{Machado:2011tn} 
  P.~A.~N.~Machado, H.~Nunokawa, F.~A.~Pereira dos Santos and R.~Zukanovich Funchal,
  ``Testing Nonstandard Neutrino Properties with a M\"ossbauer Oscillation Experiment,''
  JHEP {\bf 1111}, 136 (2011)
  [arXiv:1108.3339 [hep-ph]].
  

\bibitem{Visscher:1959}
  W.~M.~Visscher,
  ``Neutrino Detection by Resonance Absorption in Crystals at Low Temperatures,''
  Phys.\ Rev.\ {\bf 116}, 1581 (1959).





\bibitem{Daudel:1947aa}
  R.~Daudel, P.~Benoist, R.~Jacques and M.~Jean,
  ``Sur la d\'esint\'egration $\beta$,''
  Compt.\ Rend.\ Acad.\ Sci.\ (Paris) {\bf 224}, 1427 (1947),\\

\bibitem{Daudel:1947bb}
  R.~Daudel, M.~Jean and M.~Lecoin,
  ``Sur la possibilit\'e d'existence d'un type particulier de radioactivit\'e ph\'enom\`ene de cr\'eation $e$,''
  J.\ Phys.\ Radium\ {\bf 8}, 238 (1947);\\

\bibitem{Daudel:1947cc}
  R.~Daudel, M.~Jean and M.~Lecoin,
  ``Sur la d\'esint\'egration $\beta$. Ph\'enom\`ene de cr\'eation $e$,''
  Compt.\ Rend.\ Acad.\ Sci.\ (Paris) {\bf 225}, 290 (1947).  


\bibitem{Sherk:1959}
  P.~M.~Sherk,
  ``Bound Electron Creation in the Decay of Tritium,''
  Phys.\ Rev.\ {\bf 75}, 789 (1949).
  

\bibitem{Bahcall:1961zz} 
  J.~N.~Bahcall,
  ``Theory of Bound-State Beta Decay,''
  Phys.\ Rev.\  {\bf 124}, 495 (1961);

\bibitem{Bahcall:1963zza} 
  J.~N.~Bahcall,
  ``Exchange and Overlap Effects in Electron Capture and in Related Phenomena,''
  Phys.\ Rev.\  {\bf 132}, 362 (1963).



\bibitem{Takahashi:1987zz} 
  K.~Takahashi, R.~N.~Boyd, G.~J.~Mathews and K.~Yokoi,
  ``Bound-state beta decay of highly ionized atoms,''
  Phys.\ Rev.\ C {\bf 36}, 1522 (1987).


  
\bibitem{Jung:1992pw} 
  M.~Jung, F.~Bosch, K.~Beckert, H.~Eickhoff, H.~Folger, B.~Franzke, P.~Kienle and O.~Klepper {\it et al.},
  ``First observation of bound state beta-decay,''
  Phys.\ Rev.\ Lett.\  {\bf 69}, 2164 (1992).

\bibitem{Budick:1983}
  B.~Budick,
  ``Atomic Effects on the Tritium ft Value,''
  Phys.\ Rev.\ Lett.\ {\bf 51}, 1034 (1983),\\


\bibitem{Lasser:1989}
  R.~L\"asser,
  \textit{Tritium and Helium-3 in Metals}
  (Springer 1989, reprint 2012).

\bibitem{Lucas:2000}
  L.~L.~Lucas and M.~P.~Unterweger,
  ``Comprehensive Review and Critical Evaluation of the Half-Life of Tritium,''
  J.\ Res.\ Natl.\ Inst.\ Stand.\ Technol.\ {\bf 105}, 541 (2000).


\bibitem{Heisenberg:1927}
  W.~Heisenberg,
  ``\"Uber den anschaulichen Inhalt der quantentheoretischen Kinematik und Mechanik,''
  Z.\ Phys.\ {\bf 43}, 172 (1927).
  

\bibitem{Ozawa:2003}
  M.~Ozawa,
  ``Physical content of HeisenbergÕs uncertainty relation: Limitation and reformulation,''
  Phys.\ Lett.\ A {\bf 318}, 21 (2003).


\bibitem{Mandelstam:1945}
  L.~I.~Mandelstam and I.~E.~Tamm,
  ``The uncertainty relation between energy and time in nonrelativistic quantum mechanics''
  J.\ Phys.\ (USSR) {\bf 9}, 249 (1945).


\bibitem{Aharonov:1961}
  Y.~Aharonov and D.~Bohm,
  ``Time in the Quantum Theory and the Uncertainty Relation for Time and Energy,''
  Phys.\ Rev.\ {\bf 122}, 1649 (1961).

\bibitem{Aharonov:1975}
  Y.~Aharonov and J.~L.~Safko,
  ``Measurement of noncanonical variables,''
  Annals\ Phys.\ {\bf 91}, 279 (1975).

\bibitem{Anandan:1990fq} 
  J.~Anandan and Y.~Aharonov,
  ``Geometry Of Quantum Evolution,''
  Phys.\ Rev.\ Lett.\  {\bf 65}, 1697 (1990).


\bibitem{busch:1990aa}
  P.~Busch,
  ``On the energy-time uncertainty relation. Part I: Dynamical time and time indeterminacy,''
  Found.\ Phys.\ {\bf 20}, 1 (1990).

\bibitem{busch:1990bb}
  P.~Busch,
  ``On the energy-time uncertainty relation. Part II: Pragmatic time versus energy indeterminacy,''
  Found.\ Phys.\ {\bf 20}, 33 (1990).

\bibitem{busch:1990cc}
  P.~Busch,
  ``The Time-Energy Uncertainty Relation,''
  Lecture Notes in Physics {\bf 734}, {\it Time in Quantum Mechanics},
  eds. J.~G.~Muga, R.~Saka~Mayato, and \'I.~L.~Egusquiza (Springer, 2nd ed. 2007) pp.73-105
  [arXiv:quant-ph/0105049v3].





\bibitem{auletta}
For an encyclopedic review consult: 
G. Auletta, {\it Foundations and Interpretation of Quantum Mechanics : In the Light of a Critical-historical Analysis of the Problems and of a Synthesis of the Results} (World Scientific, 2000).



\bibitem{Maggiore:1993kv}
  M.~Maggiore,
  ``The Algebraic Structure Of The Generalized Uncertainty Principle,''
  Phys.\ Lett.\  B {\bf 319}, 83 (1993)
  [arXiv:hep-th/9309034].

\bibitem{Kempf:1994su}
  A.~Kempf, G.~Mangano and R.~B.~Mann,
  ``Hilbert Space Representation Of The Minimal Length Uncertainty Relation,''
  Phys.\ Rev.\  D {\bf 52}, 1108 (1995)
  [arXiv:hep-th/9412167].

\bibitem{Kempf:1996fz}
  A.~Kempf,
  ``Nonpointlike Particles in Harmonic Oscillators,''
  J.\ Phys.\ A  {\bf 30}, 2093 (1997)
  [arXiv:hep-th/9604045].

\bibitem{Chang:2001kn}
  L.~N.~Chang, D.~Minic, N.~Okamura and T.~Takeuchi,
  ``Exact solution of the harmonic oscillator in arbitrary dimensions with minimal length uncertainty relations,''
  Phys.\ Rev.\  D {\bf 65}, 125027 (2002)
  [arXiv:hep-th/0111181].

\bibitem{Chang:2001bm}
  L.~N.~Chang, D.~Minic, N.~Okamura and T.~Takeuchi,
  ``The Effect of the Minimal Length Uncertainty Relation on the Density of States and the Cosmological Constant Problem,''
  Phys.\ Rev.\  D {\bf 65}, 125028 (2002)
  [arXiv:hep-th/0201017].
  
\bibitem{Benczik:2002tt}
  S.~Benczik, L.~N.~Chang, D.~Minic, N.~Okamura, S.~Rayyan and T.~Takeuchi,
  ``Short Distance vs. Long Distance Physics: The Classical Limit of the Minimal Length Uncertainty Relation,''
  Phys.\ Rev.\  D {\bf 66}, 026003 (2002)
  [arXiv:hep-th/0204049],\\

\bibitem{Benczik:2005bh}
  S.~Benczik, L.~N.~Chang, D.~Minic and T.~Takeuchi,
  ``The hydrogen atom with minimal length,''
  Phys.\ Rev.\  A {\bf 72}, 012104 (2005)
  [arXiv:hep-th/0502222].

\bibitem{Chang:2010ir} 
  L.~N.~Chang, D.~Minic and T.~Takeuchi,
  ``Quantum Gravity, Dynamical Energy-Momentum Space and Vacuum Energy,''
  Mod.\ Phys.\ Lett.\ A {\bf 25}, 2947 (2010)
  [arXiv:1004.4220 [hep-th]].

\bibitem{Chang:2011jj} 
  L.~N.~Chang, Z.~Lewis, D.~Minic and T.~Takeuchi,
  ``On the Minimal Length Uncertainty Relation and the Foundations of String Theory,''
  Adv.\ High Energy Phys.\  {\bf 2011}, 493514 (2011)
  [arXiv:1106.0068 [hep-th]].

\bibitem{Lewis:2011fg} 
  Z.~Lewis and T.~Takeuchi,
  ``Position and Momentum Uncertainties of the Normal and Inverted Harmonic Oscillators under the Minimal Length Uncertainty Relation,''
  Phys.\ Rev.\ D {\bf 84}, 105029 (2011)
  [arXiv:1109.2680 [hep-th]].


\bibitem{Shalyt-Margolin:2002}
  A.~E.~Shalyt-Margolin and A.~Ya.~Tregubovich,
  ``Generalized uncertainty relations in a quantum theory and thermodynamics from the uniform point of view,''
  arXiv:gr-qc/0204078.



\bibitem{Padmanabhan:1987au}
  T.~Padmanabhan,
  ``Limitations on the Operational Definition of Space-time events and Quantum Gravity,''
  Class.\ Quant.\ Grav.\  {\bf 4} (1987) L107.


\bibitem{Gross:1987ar} 
  D.~J.~Gross and P.~F.~Mende,
  ``String Theory Beyond the Planck Scale,''
  Nucl.\ Phys.\ B {\bf 303}, 407 (1988).
  
\bibitem{Gross:1987kza} 
  D.~J.~Gross and P.~F.~Mende,
  ``The High-Energy Behavior of String Scattering Amplitudes,''
  Phys.\ Lett.\ B {\bf 197}, 129 (1987).

\bibitem{Amati:1988tn} 
  D.~Amati, M.~Ciafaloni and G.~Veneziano,
  ``Can Space-Time Be Probed Below the String Size?,''
  Phys.\ Lett.\ B {\bf 216}, 41 (1989).

\bibitem{Amati:1987uf} 
  D.~Amati, M.~Ciafaloni and G.~Veneziano,
  ``Classical and Quantum Gravity Effects from Planckian Energy Superstring Collisions,''
  Int.\ J.\ Mod.\ Phys.\ A {\bf 3}, 1615 (1988).
  
\bibitem{Amati:1987wq} 
  D.~Amati, M.~Ciafaloni and G.~Veneziano,
  ``Superstring Collisions at Planckian Energies,''
  Phys.\ Lett.\ B {\bf 197}, 81 (1987).

\bibitem{Witten:1997fz} 
  E.~Witten,
  ``Duality, space-time and quantum mechanics,''
  Phys.\ Today {\bf 50N5}, 28 (1997).


\bibitem{camelia}
A version of this uncertainty relation with
$\alpha=0$ and $\beta(t)=\eta\,t$ (where $\eta$ is a dimensionful constant)
has been suggested as a way to understand quantum decoherence and CPT violation:\\
  G.~Amelino-Camelia,
  ``Classicality, matter-antimatter asymmetry, and quantum gravity deformed uncertainty relations,''
  Mod.\ Phys.\ Lett.\ A {\bf 12}, 1387 (1997)
  [gr-qc/9706007].
     

\bibitem{Minic:2003en}
  D.~Minic and C.~H.~Tze,
  ``Background Independent Quantum Mechanics and Gravity,''
  Phys.\ Rev.\  D {\bf 68}, 061501 (2003)
  [arXiv:hep-th/0305193].

\bibitem{Minic:2003nx}
  D.~Minic and H.~C.~Tze,
  ``A General Theory of Quantum Relativity,''
  Phys.\ Lett.\  B {\bf 581}, 111 (2004)
  [arXiv:hep-th/0309239].

\bibitem{Minic:2004rj}
  D.~Minic and H.~C.~Tze,
  ``What is quantum theory of gravity?,''
  arXiv:hep-th/0401028.

\bibitem{Jejjala:2006jf}
  V.~Jejjala and D.~Minic,
  ``Why there is something so close to nothing: Towards a fundamental theory of the cosmological constant,''
  Int.\ J.\ Mod.\ Phys.\  A {\bf 22}, 1797 (2007)
  [arXiv:hep-th/0605105].

\bibitem{Jejjala:2007rn}
  V.~Jejjala, M.~Kavic and D.~Minic,
  ``Time and M-theory,''
  Int.\ J.\ Mod.\ Phys.\  A {\bf 22}, 3317 (2007)
  [arXiv:0706.2252 [hep-th]].

\bibitem{Jejjala:2008nc}
  V.~Jejjala, M.~Kavic, D.~Minic and C.~H.~Tze,
  ``On the Origin of Time and the Universe,''
  Int.\ J.\ Mod.\ Phys.\  A {\bf 25}, 2515 (2010)
  [arXiv:0804.3598 [hep-th]].

\bibitem{Jejjala:2009ky}
  V.~Jejjala, M.~Kavic, D.~Minic and C.~H.~Tze,
  ``The Big Bang as the Ultimate Traffic Jam,''
  Int.\ J.\ Mod.\ Phys.\  D {\bf 18}, 2257 (2009)
  [arXiv:0905.2992 [gr-qc]].



\bibitem{penrose}
For this suggestion consult:
R.~Penrose, \textit{The Road to Reality: A Complete Guide to the Laws of the Universe} 
(Knopf 2005, Vintage 2007).\\
For more recent references see Refs.~\cite{'tHooft:2007xi} and \cite{Weinberg:2011jg},
and references therein.\\


\bibitem{'tHooft:2007xi} 
  G.~'t Hooft,
  ``Emergent Quantum Mechanics and Emergent Symmetries,''
  AIP Conf.\ Proc.\  {\bf 957}, 154 (2007)
  [arXiv:0707.4568 [hep-th]].

\bibitem{Weinberg:2011jg} 
  S.~Weinberg,
  ``Collapse of the State Vector,''
  Phys.\ Rev.\ A {\bf 85}, 062116 (2012)
  [arXiv:1109.6462 [quant-ph]].



\bibitem{Chang:2011yt}
  L.~N.~Chang, Z.~Lewis, D.~Minic, T.~Takeuchi, and C.~H.~Tze,
  ``Bell's Inequalities, Superquantum Correlations, and String Theory,''
  Advances in High Energy Physics {\bf 2011} (2011) 593423
  [arXiv:1104.3359 [quant-ph]].


\bibitem{weinberg:cosmo}
See for example: S. Weinberg, \textit{Cosmology} (Oxford University Press, 2008).

\end{thebibliography}
\end{document}